# From MANET to people-centric networking: milestones and open research challenges

Marco Conti, DIITET-CNR, Italy
Chiara Boldrini, IIT-CNR, Italy
Salil Kanhere, University of New South Wales, Australia
Enzo Mingozzi, Università di Pisa, Pisa, Italy
Elena Pagani, Università degli Studi di Milano, Italy
Pedro M. Ruiz, Universidad de Murcia, Spain
Mohamed Younis, University of Maryland, Baltimore County (UMBC), USA

*Abstract*

In this paper we discuss the state of the art of (mobile) multi-hop ad hoc networking with the aim to present the current status of the research activities and identify the consolidated research areas, with limited research opportunities, and the hot and emerging research areas for which further research is required. We start by briefly discussing the MANET paradigm, and why the research on MANET protocols is now a cold research topic. Then we analyze the active research areas. Specifically, after discussing the wireless-network technologies we analyze four successful ad hoc networking paradigms, mesh, opportunistic, vehicular networks, and sensor networks that emerged from the MANET world. We also present the new research directions in the multi-hop ad hoc networking field: people centric networking, triggered by the increasing penetration of the smartphones in everyday life, which is generating a people-centric revolution in computing and communications.

**1 Introduction**

This article discusses the state of the art of (mobile) multi-hop ad hoc networking [1.1]. The aim is *i)* to point out research topics that have been extensively researched and are now cold research areas, for which further research is not recommended, and *ii)* to identify the hot and emerging research challenges that still require extensive investigations.

The multi-hop (mobile) ad hoc networking paradigm emerged, in the civilian field, around 90's with the availability of off-the-shelf wireless technologies able to provide direct network connections among users devices: Bluetooth (IEEE 802.15.1), for personal area networks, and the 802.11 standards' family for high-speed wireless LAN (see Chapters 2-4 in [1.2]). Specifically, these wireless standards allow direct communications among network devices within the transmission range of their wireless interfaces thus making single-hop ad hoc network a reality, i.e., infrastructure-less WLAN/WPAN where devices communicate without the need of any network infrastructure.

The multi-hop paradigm was then conceived to extend the possibility to communicate to any couple of network nodes, without the need to develop any ubiquitous network infrastructure: nearby users directly communicate (by exploiting the wireless-network interfaces of their devices in ad hoc mode) not only to exchange their own data but also to relay the traffic of other network nodes that cannot directly communicate, thus operating as the routers do in the legacy Internet [1.3]. This paradigm has been often identified with the solutions developed inside the MANET IETF working group, and for this reason was often called the MANET paradigm [1.4].

At its born, MANET was seen as one of the most innovative and challenging wireless networking paradigm and was promising to become one of the major technologies, increasingly present in everyday life of everybody. Therefore, MANET immediately gained momentum and this produced tremendous research efforts in the mobile-network community [1.3]. MANET research focused on what we call *pure general-purpose* MANET, where *pure* indicates that no infrastructure is assumed to implement the network functions and no authority is in charge of managing and controlling the network. *General-purpose* denotes that these networks are not designed with any specific application in mind, but rather to support any legacy TCP/IP application. Following this view, the research focused on enhancing and extending the IP-layer routing and forwarding functionalities in order to support the legacy Internet services in a network without any infrastructure. At network layer, we observed a proliferation of routing protocol proposals, as legacy Internet routing protocols developed for wired networks are clearly not suitable for the unpredictable and dynamic nature of MANET topology. The research interests rapidly spread from routing to all layers of the Internet protocol stack: from the physical up to the application layer [1.5]. Applications, probably constitutes the less investigated area of MANETs. Indeed, in the design of general-purpose MANETs, there was not a clear understanding of the applications for which multi-hop ad hoc networks are an opportunity. Lack of attention to the applications, probably, represents one of the major causes for the negligible MANET impact in the wireless networking field.

After about two decades of intense research efforts, the MANET research field produced profound theoretical results (e.g., performance bounds on MANET performance [1.6][1.7]), or innovative protocols and architectural solutions (e.g., cross-layer architectures and protocols) [1.5], but in terms of real world implementations and industrial deployments, the pure general-purpose MANET paradigm suffers from scarce exploitation and low interest in the industry and among the users [1.4]. Therefore, in the last years, the interest for the MANET paradigm has been rapidly decreasing and the research on MANET protocol can now be considered a cold research topic. Indeed it is interesting to observe that the number of manuscripts focusing on MANET, published in top quality journals, is rapidly decreasing.[1]

Even though MANET research has not a major impact on the wireless networking field (except for specialized fields such as military and disaster recovery), several networking paradigms have emerged from the MANET field that are currently generating interest both in the academia and industry. Among these, let's remember: mesh networks, vehicular networks, opportunistic networks and sensor networks. These multi-hop ad hoc networking paradigms, by learning from the MANET experience, emerged from MANET by avoiding the main weaknesses of the MANET research by following a more *pragmatic development strategy* [1.4].

In this article, after a review of the current status of wireless technologies for multi-hop networks (see Section 2), we discuss the open research challenges in mobile ad hoc networking paradigms: mesh networks (Section 3), opportunistic networking (see Section 4), vehicular networks (see Section 5) and sensor networks (see Section 6). Section 7 is devoted to the emerging people-centric paradigm that, thanks to the increasing diffusion of the smartphones, combines wireless communications and sensor networks to build computing and communication solutions that are tightly coupled with the daily life and behaviors of people.

## 2. Technologies for Mobile Multi-hop Wireless Networks[2]

Many technologies and standards, enabling multi-hop wireless networking as well as off-the-shelf equipment implementing such technologies, are available today [2.1]. Nevertheless, we can easily foresee that research interest in these topics will not diminish in the next years, but rather it is likely to grow. There are at least two reasons for this.

On the one hand, the development and availability of technologies and equipment is driven by standardization and regulation activities. Networking standards, thus including multi-hop wireless ones, are primarily concerned with ensuring device interoperability, a key requirement to allow for networking equipment mass production and market diversification. Nevertheless, guaranteeing interoperability does not require a full system specification; rather, standards usually do not specify key aspects that do not affect interoperability to create additional space for competition among manufacturers. One such example is the resource allocation/scheduling on the data plane, whose implementation strongly affects performance but does not concern interoperability. Therefore, new technologies and standards, even though leaning upon a layer of fundamental research results, always call for investigating many practical aspects that are mainly related to system customization and performance optimization in concrete scenarios. In the following sections, we will highlight the challenges that we envision for a set of relevant technologies and standards related to wireless multi-hop networks.

On the other hand, one key advantage of mobile multi-hop/mesh wireless communication technologies is that they allow for easy and fast, highly scalable and cost-effective network deployment under heterogeneous environments. Therefore, there are many heterogeneous application scenarios where multi-hop wireless technologies represent one of the most effective solutions, if not the only viable one. Many of such scenarios are real today, like public city-wide dedicated services (e.g., video-surveillance or transport systems), Internet access in rural areas, environmental monitoring through sensing systems, disaster emergencies, etc. As observed in [1.4], a pragmatic development approach allowed a bunch of research activities to thrive focusing on the specific challenges and requirements of these different scenarios in order to build robust and effective networking solutions (either mesh, sensor, opportunistic, or vehicular). We think however that new application and deployment scenarios, for which multi-hop wireless networks can play a crucial role, will just grow exponentially in the near future, as the *Internet of Things* vision will become a reality [2.2]. Connecting everything, anytime and everywhere will demand for capillary and pervasive network coverage, for which multi-hop wireless technologies represent an effective solution. However, such scenarios will pose many new challenges – one on top of all others: scalability – which will entail reconsidering current results and will likely boost a new wave of research on multi-hop wireless technologies. Let us briefly review some of them in the following.

---

[1] For example, since 2013, only one paper with the keyword "MANET", in the title, has been published in *Computer Communications*. Furthermore, only a few papers, focusing on MANET issues, have been published in this journal in the last three years.
[2] By Enzo Mingozzi

**Smart Cities**. This term identifies a generic scenario where *Information and Communication Technologies* are employed to sustain the development of urban environments. This is achieved by the pervasive deployment of sensing, computation, and communication infrastructures in order to collect and distribute the information needed to develop and improve new services offered to citizens by municipalities. As such, this scenario is indeed comprehensive of many heterogeneous ones. In most cases, the communication infrastructure is characterized by frequent topology changes, faults in equipment, and/or harsh environmental conditions, which naturally call for the use of multi-hop wireless network technologies. All scenarios, however, share some common and unique features, e.g., very dense networks comprising thousands of devices, and uncommon traffic patterns made of a very large number of low data-rate flows. As highlighted in [2.3], this needs reconsidering traditional network design and technologies, and open new challenges in terms of standardization (including multi-hop communication), cooperation (e.g., managing interference in dense environments), security, QoS [2.4], etc. Research activities addressing these new challenges are already ongoing, and more are expected to come, as the Smart City scenario will evolve in the future.

**Industrial Internet**. Communication networks are a key component of infrastructures supporting complex industrial processes. Being involved in the control process, they have very critical requirements in terms of reliability and QoS [2.5]. To meet such requirements, industrial networks typically use wired connections, and dedicated standards have been developed. On the other hand, the need to reduce costs is triggering a new trend towards IP-based solutions that leverage multi-hop wireless technologies allowing for cost-effective industrial network deployments [2.6]. However, providing *ultra-high* reliability and *deterministic* QoS guarantees in multi-hop wireless networks is a challenging task, combined with the need of implementing low-power consumption solutions. Though already investigated in the past, these issues are still open and require new research work specifically focusing on the Industrial Internet scenario.

**Machine-to-Machine applications (M2M)**. Though partially overlapping with other scenarios, this one is characterized by the deployment of communication technologies aiming to interconnect devices (machines) involved in control and monitoring applications that do not require humans in the loop. The reason for having many of these applications is to remove human intervention for improved efficiency, security, and safety. As such, industrial applications are an instance of M2M applications, but they are not the only scenario of interest. In fact, when referring to M2M communications, ubiquitous connectivity is usually considered a mandatory requirement, which is not usually the case for industrial deployments. Therefore, mobile broadband wireless network technologies are mainly in the spotlight [2.7]. Requirements are QoS provisioning, energy-efficiency and co-existence with other communication types, e.g., Human-to-Human (H2H), under specific traffic patterns [2.8] and very large density of devices. This is challenging to achieve in mobile cellular networks, and needs considering novel approaches such as Heterogeneous Networks (HetNets for short) [2.9] and Device-to-Device (D2D) communication [2.10]. In both cases, multi-hop relaying provides benefits that will be illustrated in more detail in the next sections.

**Smart grid**. It is today commonly recognized that upgrading current power grids operation by integrating advanced sensing, communication and control functionalities is mandatory for the purpose of improving reliability and safety, increasing efficiency by cost reduction and reducing $CO_2$ emissions [2.11]. This requires the design and deployment of an integrated communication network including different segments in terms of geographic scale and functionality, for which therefore different communication technologies are applicable. These networks are usually referred as Home Area Networks (HANs), covering a single residential unit, Neighborhood Area Networks (NANs), including a cluster of homes usually powered by the same transformer, and Wide Area Networks (WAN), which connect NANs to the utility operator. Moreover, equipment in the field requires to be controlled by a separate network called Field Area Network (FAN) having a geographic scale similar to NANs. For all these different networks, multi-hop wireless technologies, either broadband, local area, or short range, depending on the coverage requirements, are an effective solution. However, new Smart Grid specific applications are planned to be realized through the Smart Grid network infrastructure, like Advanced Metering, Demand Response, Wide Area Situational Awareness, and Distributed Energy resources and storage. Based on the data requirements of these applications, new challenges have arisen at the network layer of the protocol stack [2.12]. In particular, these applications are expected to have high security, high reliability, and various QoS requirements such as bandwidth and latency, which need to be addressed by routing and data forwarding in the first place [2.13][[2.14], and likely then through cross-layer approaches. Energy efficiency is also a key requirement when low-power devices are used [2.15]. For HANs and NANs, the standardization activity carried out by the IETF ROLL (Routing Over Low power and Lossy networks) working group is also relevant.

**Personal healthcare**. There has been a tremendous development and advancement of micro- and nano-technologies that allow building miniaturized low-power sensors with (wireless) communication and computation capabilities that can be placed on a human body for various medical (e.g. personal health monitoring) and even non-medical (e.g., personal entertainment) applications [2.16]. When connected together, they form a Wireless Body Area Network (WBAN), which aim to provide fast and reliable communication among sensors/actuators within and on the human body, and with the outside networks in proximity.

The above scenarios are certainly not the only relevant ones for wireless multi-hop technologies. In **Smart Homes**, for example, wireless multi-hop communication is involved both in wireless LAN technologies delivering high-definition multimedia streams to multiple terminals and in wireless sensor and actuator networks for home automation. On the other hand, **Intelligent Transportation Systems** require networking infrastructures for vehicle to roadside communications that involve wireless multi-hop communications, like, e.g., in wireless mesh networks supporting node mobility and efficient handoffs.

Future research activities on multi-hop wireless networks will have to face the many challenges described above depending on the specific scenario considered. Besides those challenges, however, it is of paramount importance that performance evaluation results will be credible enough to provide real insights in the studied technologies. To this aim, it is an integral part of research efforts to produce realistic simulation models for such diverse scenarios, and to develop real testbeds to evaluate experimentally, with the help of users, the technologies under consideration.

Without claiming to be comprehensive, in the following sections we delve into some of the current and envisioned technologies and standards mentioned above in order to provide more details on specific requirements and identify the most important challenges that need to be addressed. For the ease of reading, they are grouped into three classes depending on their typical coverage: mobile broadband access and cellular networks, local area networks, and, finally, low-power and short range networks.

## 2.1 Mobile Broadband Wireless Access

Multi-hop communications Broadband Wireless Access (BWA) networks have been historically neglected since central coordination makes multi-hop an inherent complex task. However, the very challenging requirements of ITU IMT-Advanced systems finally required to deal with such complexity in order to gain in terms of spectral and energy efficiency. Relay support was first introduced in IEEE 802.16m (Mobile WiMAX) [2.17] and LTE-Advanced (Rel. 10). Although WiMAX was developed before LTE as a 4G technology, interest on it has rapidly declined as the carrier market has definitely opted for the latter. In the following, we therefore focus on LTE/LTE-A and specific research challenges related to multi-hop communication.

**Heterogeneous Networks (HetNets)** refer to the heterogeneous deployment of low-power nodes within an LTE/LTE-A macrocell with the aim of boosting overall spectral and energy efficiency by means of enhanced spatial resource reuse [2.18][2.19]. One such node may serve a *small cell* underlying the macrocell network (e.g., a picocell or femtocell) or act as a Relay Node (RN) between User Equipments (UEs) and the serving evolved NodeB (eNodeB, known as the donor eNB, or DeNB). In the latter case, the backhaul link is wireless, thus forming a multi-hop wireless network. RN deployment in HetNets have benefits in terms of improved cell edge performance and load balancing (typically, by deploying relays at the edge of the macrocell), reduced power consumption, and increased capacity. However, it also brings several challenges to radio resource management in terms of interference mitigation, radio resource utilization, QoS and implementation complexity. A comprehensive survey of state-of-art radio resource management schemes for LTE/LTE-A multi-hop networks, and related open research challenges, can be found in [2.20].

**Device-to-Device (D2D) communication** is a network paradigm that allows direct communication between devices in an infrastructure-based wireless network without the involvement of a central access point/base station. While largely exploited in wireless local and personal area networks, D2D communication has been considered recently also for cellular network technologies as a further means to improve spectral efficiency [2.10]. In a cellular network, D2D is mainly intended for single hop communication between UEs, and it is centrally coordinated by a supervising entity (e.g., eNB), which is finally responsible for radio resource management. However, D2D communication also enables multi-hop relay of data, i.e., a UE may act as a relay between another UE and the base station, thus enlarging cellular coverage and potentially improving the performance of cell edge users, without incurring the cost of deploying conventional relay nodes as with HetNets [2.21][2.22]. D2D communication may also provide an efficient solution for M2M communications in cellular networks [2.23], whereby the UE is used to relay M2M data from multiple machines to the base station. This allows a cell to better scale with the high control overhead and possible congestion due to the very large number of low data rate flows of a typical M2M communication system. Research challenges for D2D communication focus on different aspects depending on how the cellular spectrum is allocated to D2D links. With *underlay inband* D2D the spectrum is shared, therefore interference mitigation is of utmost importance. On the other hand, with *overlay inband* D2D dedicated resources are allocated to D2D links though the spectrum is still shared with cellular links: optimizing radio resource allocation is more relevant in this case. Finally, with *outband* D2D architectural issues due to the use of multiple radio interfaces and heterogeneous wireless technologies need to be faced [2.10]. When D2D communication is used for multi-hop relay, a key challenge is how to properly select a relay node for optimizing a performance measure of interest (e.g., maximize the capacity). The relay being a user device, it is also of interest to investigate mechanisms for motivating users to allow data relaying through their own UEs, considering that this implies increased battery draining as well as reduced available capacity. D2D communication is currently considered by 3GPP for standardization; however, it is limited to the use case of Proximity Services in Public Safety scenarios, where multi-hop data relaying is also relevant [2.24].

## 2.2 Local area networks

Enabling multi-hop communication in wireless LANs has received a lot of interest in both academia and industry since the very early days when WLAN technologies and standards were developed. This is mainly because WLAN technologies are typically operated in license-free bands and therefore have limited coverage by definition; multi-hop communication is hence the only viable option to overcome such limitations. Due to its widespread diffusion, the IEEE 802.11 standard dominates in this area as a key technology enabler for multi-hop networks of various kinds, including MANETs and Wireless Mesh Networks (WMNs) [2.1][2.25]. From a standardization point of view, the IEEE 802.11 standard did not originally include support to multi-hop transmissions, and it took a long time to extend it to provide mesh networking support as a result of the work of Task Group "s" inside the 802.11 Working Group. However, the 802.11s amendment has failed to meet industry acceptance, and available commercial solutions are typically based on proprietary extensions. More recently, the **IEEE 802.11ah** TG has been chartered to meet the typical requirements of IoT scenarios, where thousands of devices generating low-data rate flows need to be connected in the same area by an energy-efficient protocol (which is not the case of IEEE 802.11) [2.26]. Multi-hop communication is considered by 802.11ah to increase coverage and reduce energy consumption: infrastructure BSS is in fact extended with two-hop relaying, which is expected to be much easier to implement than the 802.11s framework. However, the definition of the forwarding mechanisms as well as of the modifications needed at the MAC layer is still at an early stage.

## 2.3 Low power/Short range networks

**Low-power and Lossy Networks** (LLNs) refer generally to networks consisting of a very large number of constrained nodes (i.e., with limited processing power, memory, and energy when battery operated) operating over lossy links, that typically support only low data rates and are usually unstable. Though LLN links can also be wired, LLNs based on wireless mesh networks is the most common case, leveraging the IEEE 802.15.4 low-rate WPAN standard. LLNs are the typical instance of capillary networks (i.e., edge networks) in a large class of IoT scenarios. To meet the challenging IoT requirements in terms of interoperability, diversity of applications, and scalability, protocols at different layers have been specified by IETF in order to enable IPv6 over LLNs, including the 6LoWPAN adaptation layer and the RPL routing protocol [2.27][2.28]. RPL is a distance-vector based protocol allowing for multiple instances running concurrently which optimize multi-path forwarding to specific destinations according to different objective functions depending on link costs and node attributes/status information. This is needed to meet the diverse routing requirements from heterogeneous applications running on top of the same LLN. Although some basic objective functions are specified by IETF, investigating what combination of metrics/constraints and parameter configuration is optimal depending on the use case requirements remains an open issue [2.29].

More recently, the **IEEE 802.15.4e** amendment has introduced a new TDMA-based MAC protocol named TSCH (Time-Slotted Channel Hopping), to facilitate the development of LLNs with a more deterministic behavior in terms of delay and power consumption, thus allowing to meet the high reliability requirements of the Industrial Internet [2.6]. In particular, TSCH allows for the advanced scheduling of transmission patterns in different time slots across the whole (wireless mesh) network. However, as expected, it does not specify any mandatory algorithm to determine and maintain such schedule, depending on the multi-hop network topology and characteristics, and the traffic requirements. The use of TSCH is therefore subject to many possible optimizations and customizations that are still open issues. The **IETF 6TiSCH** working group is currently working on specifying an architecture to enable IPv6 over TSCH and integrate it with 6LoWPAN and RPL.

**Wireless Body Area Networks** (WBANs) leverage the IEEE 802.15.6 standard for low power, short range and extremely reliable wireless communications supporting a large range of data rates within the surrounding area of the human body [2.30]. Research challenges involve all layers of the protocol stack. At the PHY layer, channel modeling and antenna design very much depend on the challenging environment (the human body) [2.16]. Meeting power requirements with diverse traffic characteristics while providing extremely high reliable throughput and delay is of utmost importance in this context and remains an open challenge at the MAC layer [2.31][2.32]. In particular, deterministic performance guaranteed by TDMA may not be suitable since periodic synchronization requires extra energy consumption, and viable alternatives need be investigated [2.33]. When multi-hop communication is adopted, developing suitable routing protocols remains a key challenge, given the requirement of maximizing network lifetime and considering that frequent topology changes occur due to body movements [2.16].

Radio frequency (RF) energy transfer and harvesting, i.e., the capability of converting the received RF signals into electrical energy, is an emerging technology that has recently received a lot of attention as an alternative proactive method to power nodes in a wireless network. In **RF energy harvesting networks**, a key challenge is to develop network functionalities that are energy-aware in the traditional sense, i.e., power consumption is reduced, but also, and most importantly, with the new meaning of using the network also for energy transfer and harvesting besides data communication. At the MAC layer, this entails coordinating channel access so as to reserve resources (i.e., time) for RF energy harvesting. When multi-hop communication is considered, routing protocols also need to be specifically designed to ensure a proper energy distribution to maintain end-to-end communication [2.34].

## 3. Wireless Mesh Networks[3]

The design of the Wireless Mesh Network (WMN) paradigm started from a well-defined set of application scenarios that can be summarized as in [3.1]: "providing a flexible and low cost extension of the Internet". Starting from the well-defined application scenario, the WMN paradigm introduces an architectural shift, with respect to MANET, by adopting a two-tier network architecture based on multi-hop communications. A multi-hop wireless backbone is formed by dedicated (and often fixed) wireless mesh routers, which run a multi-hop routing strategy to interconnect with each other. Furthermore, some mesh routers act as gateways, providing the WMN with a direct connection to the Internet and other wired/wireless networks. Finally, to support seamless data transport services for users' devices (also called mesh clients), mesh access points are connected to the mesh routers to offer connectivity to mesh clients. Consequently, topology changes due to mobility or energy issues do not influence traffic forwarding in WMNs as in MANETs, and the impact of the mobility is restricted to the last hop, i.e., the connection between the users and the mesh access points [3.1].

According to the two-tier network architecture, WMNs research activities have been subdivided in three main areas:

i) Exploiting the wireless mesh routers to build a robust network backbone interconnecting the mesh routers, and possibly some gateways to/from the Internet;

ii) Defining a set of routing protocols which, by exploiting the network backbone can identify the best path(s) for traffic forwarding inside the WMN and to/from the Internet;

iii) Supporting the users' mobility among mesh access points;

All areas have been subject of intense research activities providing effective solutions for these problems. In particular, a large body of research focused on building a robust wireless mesh backbone by exploiting advanced multi-radio, multi-channel and multi-rate capabilities, using heterogeneous wireless technologies and directional antennas [3.2]. This means that a WMN environment is characterized by a rich link and path diversity, which provides an unprecedented opportunity to find paths that can satisfy the application requirements. Effective channel assignment solutions have been developed to construct robust and efficient multi-path wireless-mesh backbones [3.3][3.4], while several routing protocols able to exploit the link and path diversity have been designed, e.g., [3.5][3.6][3.7]. The multifaceted characteristics of a WMN complicate the path selection process. For example, the interference in WMNs is a very complex phenomenon, and the routing process must be aware of the interference existing between links and traffic flows to take advantage of the multi-channel and multi-radio capabilities [3.8]. In addition, in most application scenarios the portion of traffic that a mesh router delivers to other routers in the network (i.e., intra-mesh traffic) will be minimal with respect to the traffic conveyed over connections established with external hosts (i.e., inter-mesh traffic). As a result, most WMN traffic is usually between the mesh clients and the mesh gateway(s). This implies that traffic gets aggregated at the mesh gateways, which can have a negative effect on the network performance. For instance, depending on the network topology and the routing strategy, many mesh routers may select the same gateway, and congestion levels can build up excessively on the shared wireless channel around the gateway. This may also lead to an uneven utilization of the gateways' resources. As a consequence routing protocols must carefully take into consideration the distribution of traffic loads (e.g., see "Quality of Service in Mesh Networks" in [1.1]).

Nowadays, WMN is a consolidated networking paradigm to realize networking backbone infrastructures in a cost-effective manner for different purposes, e.g., providing public Internet access in metropolitan areas[4] (supporting a wide range of services ranging from security surveillance to intelligent transportation services), or supporting connectivity in in harsh environments where cabling is not an option. From a research standpoint, the interest in WMN has been progressively diminishing but some aspects are still active research areas, such as: efficient mobility support (e.g., for vehicle-roadside communication in urban transport systems at medium-to-high speeds) [3.9], advanced Quality of Service support at very high data rates (e.g., for multimedia distribution in smart homes) [1.4] and robust, secure, and effective routing and forwarding protocols, e.g., [3.10]-[3.13].

---

[3] By Marco Conti
[4] See http://www.muniwireless.com/

# 4. Opportunistic networking[5]

Opportunistic Networks (ONs) are an evolution of Mobile Ad Hoc Networks (MANETs), and are considered as a particular case of delay/disruption tolerant networks (DTNs). While in MANETs mobility is assumed such that end-to-end paths exist between source-destination pairs, and path disruption is a possible but not usual event, in ONs the network is frequently partitioned, due to battery discharge, to the short radio range combined with both mobility and node sparseness, and to the low reliability of wireless links. For this reason, source-destination paths are likely to never exist. Thus, while MANETs can *hide* node mobility, ONs need to *expose* node mobility and exploit it through the store-carry-and-forward paradigm [4.1]. According to this paradigm, data are "moved" across the network not only through message passing amongst nodes, but also through the movements of the nodes themselves, which carry messages while waiting to enter in reciprocal radio range with either a message destination or a node suitable to reach the destination. In this sense, ONs are similar to DTNs, although the standard bundle protocol [4.2] designed for DTNs seems inadequate for ONs. Opportunistic networking has been traditionally considered as a mean to provide communication services mainly to portable devices (e.g., smartphones) in the absence of a fixed infrastructure, or when the cellular network is considered not convenient in terms of either bandwidth availability or cost. It is in this form that they have been extensively studied in the literature, as we discuss in Section 4.1. However, as we show later on, there are challenging and exciting recent developments that promise to bring opportunistic networks research into a new and fascinating era.

## 4.1 ONs: enabling mechanisms and consolidated research results

At its onset, research on opportunistic networks focused on the *enabling mechanisms* for opportunistic data exchange, i.e., i) on understanding human mobility and human social relationships in order to exploit them when forwarding messages, and ii) on the basics of data delivery in standalone ONs.

The **dynamics of encounters** between nodes is of paramount importance, as their duration affects the amount of data that can be exchanged between two nodes, while their frequency and pattern affect delivery delay and delivery probability, respectively. This is why the gathering of mobility and encounter data, their analysis and modelling, are one of the most investigated fields. Large public repositories of encounter and mobility traces are available for researchers (e.g. CRAWDAD [4.3]), and many works have been published that analyse those traces [4.4]-[4.11] to cite just a few). It is now well recognized that several parameters describing human mobility and encounters follow a power law distribution, such as the distance covered by a movement between two places; the visiting time, return time, and visiting frequency of a location; the contact duration and inter-contact time between pairs of persons (see, e.g., [4.12] for an extensive survey). Based on these results, several mobility models have been implemented [4.13][4.14][4.15][4.16] amongst others), able to supply synthetic traces satisfying the power law distribution and well approximating real traces, which can be used for experimentation and validation of solutions designed for ONs.

Data delivery in ONs takes the form of **routing and content dissemination**. In routing, communications are host-centric, since messages have explicit source and destination. In data dissemination, communications are content-centric: messages are tagged based on their content and the goal is to bring messages to users that are interested in this content regardless of the identities of content consumers/producers. Routing issues in ONs have been thoroughly analysed in the literature, from both an algorithmic and a performance modelling standpoint (see [4.17][4.18] for a survey of the main results in this area). Since randomised routing schemes (e.g., [4.19][4.20][4.21]) have proven valuable only for quite homogenous mobility environments, the spotlight in the routing scene has been taken by utility-based schemes. Utility-based schemes ([4.22]-[4.27] to cite a few) are able to exploit the unique forwarding features of individual nodes in terms of frequency of contacts, social relationships, general context, and even resource availability (memory, battery, etc.) on the user device. Alongside algorithms, recently several analytical models have also been proposed, ranging from models that address simple homogeneous mobility scenarios to models that address heterogeneity and also depart from the simplifying exponential assumption for inter-contact times [4.28]-[4.31]. Due to the availability of this huge body of work, recent proposals for new ONs routing protocols often represent minor variations on the main ideas behind the established protocols mentioned above. Thus, we argue that — unless a paradigm shift is proposed or new evidence emerges from field experiments or new applications are envisioned that call for improved or refined routing strategies (see e.g. opportunistic crowd sensing discussed later on) — *research activities on this topic should be put on hold*.

---

[5]   By Chiara Boldrini and Elena Pagani

Content dissemination in ONs was also quite exhaustively investigated in the past years. Differently from routing, in this case a node stores a content because either the local user is interested in it, or the node is willing to cooperate in the content dissemination when it encounters interested nodes (possibly, it is the content generator). The goal here is to select what content items a node should store, as a trade-off of several local and global aspects, such as local memory availability and interest in the item, probability for the node to encounter other nodes interested in the item, item popularity, number of item copies in the system (i.e., of nodes concurrently performing that item dissemination), load balancing amongst nodes. In most proposals (e.g., [4.32][4.34][4.35]), these parameters are weighted through utility functions that rank the items so as to balance effective delivery vs. resource consumption. The ranking prioritizes the exchange of items most relevant and less disseminated so far. Alternatively, the items may be related to a certain geographical area within which their spreading is bounded [4.33]. In all cases, the decision must be taken based only on local observations. Works exist that consider the issue of performing accurate decisions in spite of maintaining incomplete observation history [4.36]. The possibility of breaking up large content into pieces and deciding what pieces to exchange has also been investigated [4.37]. Open areas of research in this field concern *the use of recommender systems and machine learning mechanisms to predict which content items may be of interest for which users* thus providing indications to caching policies. Adopting those mechanisms could help to incentivize the cooperation amongst users, thus supplying better performance in terms of both user satisfaction in obtaining content of interest, and convenient resource usage.

### 4.2. ONs out of the research lab

While research on standalone opportunistic networks has thrived, it is now time for ONs to exit the research realm and enter into people's real lives. However, two factors act as the main showstoppers: the human factor and the technology factor. As for the **human factor**, performing experiments of significant size and temporal length requires the repeated involvement of a user base willing to devote their time and patience to testing the prototype of a research product. While this can still be feasible for a one-time experiment (e.g. recruiting people attending a conference for several days) it is hard to replicate the experiment later on or bring it in a real daily life environment. However, this is crucial for testing the effectiveness of opportunistic networks and how people might react to them. In fact, *the way in which people are actually inclined to collaborate and share the resources of their devices* for the goals of the ONs *remains not yet understood*. Indeed, when two nodes come in reciprocal radio range, there are several reasons why they might decide to *not* exploit the formed link, e.g. resource shortage (in terms of battery power, computation or memory availability), scarce trust in the other part, selfishness, lack of interest in the content the other node would like to relay. This attitude can impact on the real connectivity of the ON and thus on their communication capability, and might require the development of adequate incentive policies. Several studies have been carried out to understand the effect of human behaviour on the performance of ONs [4.38]-[4.40] and solutions have been proposed to overcome selfishness [4.41] and to incentivise cooperation [4.42]-[4.44]. However, all these works are based on some abstract and general model of selfishness with little ground truth specific to ONs. To the best of our knowledge the only empirical evaluation of altruism in opportunistic networks has been provided in [4.45].

A better insight into the above issues can be gained only through the **deployment of prototypes and applications** whose widespread use would provide significant data about how people use ONs. Unfortunately, very few applications for opportunistic networking are available, which can be classified into four categories. First, there are prototypes developed for specific research purposes, such as gathering data about opportunistic contacts [4.46]. To this class belong also proofs of concepts such as the opportunistic jukebox in [4.47], where users opportunistically exchange musical content among themselves and/or with fixed points in the environment, and Ateneo On Fly [4.48], which allows content sharing among professors and students in a campus, where infostations are adopted to mediate the communications. Second, and perhaps similar, there are experimental opportunistic infrastructures to supply networking services in rural areas, such as [4.49], and [4.50], which supports search of both friends and data, and resource sharing. The applications in both classes above are likely of extemporary or niche use, and they are not available to consumers. Third, there are middleware infrastructures that do not provide applications per se; rather they supply services to deploy applications over an ON. An example is the CAMEO platform [4.51], which maintains the context of each user, allowing to make users aware of their neighbours and of the existence of content that may be of interest for them. Finally, there are genuine apps available in stores; two examples are Twimight [4.52] and SocialPal [4.53]. Twimight is an app for Android devices that allows to epidemically spread a tweet amongst users having the app installed, exploiting opportunistic encounters and the Bluetooth technology. Messages may be sent to the Internet once a participant device detects an Internet connection available. It has been designed to work in disaster scenarios, where cellular networks become unavailable. The SocialPal framework includes applications enabling communication through opportunistic interactions. For instance, the nearbyPeople app allows a person to discover the common friends in online social networks shared with users in the neighbourhood, without revealing his/her own identity, thus supporting the establishment of new relationships. So far, all the above applications are used by small communities of users, especially in comparison with the huge base of potential users. Hence, *there is still the need of large scale test beds that provide insights on the use of ONs in real environments and of widespread commercial apps that make ONs popular among users.*

It is interesting to notice that all the above prototypes rely on one-hop ad hoc data exchange. This is a direct consequence of the difficulty in using existing wireless technologies to implement opportunistic communications, what we have referred to as the **technology hindrance**. In fact, there are *no mature solutions implemented on off-the-shelf mobile devices that enable seamless multi-hop ad hoc communications, and even currently available research solutions (e.g. WiFi-Opp [4.54]) are suboptimal*. On Android devices, WiFi Direct, which is considered the best viable alternative by many, was not designed with opportunistic networking in mind and often requires cumbersome configuration (in particular when multiple groups of devices are involved [4.55]) and annoys the user with many interruptions for explicit connection acceptance. Opportunistic communications on iOS devices are even less understood. We believe that this is a major obstacle to the deployment of opportunistic networks in real life, thus representing one of the top priorities from the research standpoint.

Even playing with the technological cards we are dealt, there still remains the problem that *ad hoc communications tend to be very energy hungry [4.56]*. This affects both the QoS in the network (because devices with no battery left simply will not deliver messages on behalf of other nodes) and user satisfaction (users will be less willing to join an ON if they are afraid their battery will be drained in a few hours). There are not many works in the literature explicitly addressing the problem of **energy optimisation** in ONs. Those who do, mainly pursue the two lines of (i) optimising neighbour discovery (which is typically very expensive in terms of energy) [4.57][4.58][4.59] and (ii) alternatively switching roles when a different energy expenditure is involved in the host-client roles of a pairwise wireless connection [4.60]. Despite these attempts, energy issues in ONs still remain mostly unaddressed, and an open research challenge for ONs of the future.

### 4.3. Crowd sensing/computing and Opportunistic IoT

According to the crowd-sensing paradigm, people with their smart devices, willingly or unwillingly, represent potential sensing devices distributed across cities, see Section 7. For example, the microphones of smartphones can be used to measure the noise level in the environment, and these samples, put together, can provide a very detailed map of the noise pollution across a reference area. Given the distributed and spontaneous nature of this sensing paradigm, the use of opportunistic communications has been recently put forward. However, as discussed in [4.61], data delivery solutions designed for legacy opportunistic networks can unlikely be taken as they are, since they *fail to take into account the specificity of a crowd sensing environment*, such as the spatio-temporal correlation between sensory data. In addition, sensory data are typically of small size but manifold, so the network might be overloaded very soon. For this reason, in-network processing techniques should also be exploited.

Besides for reducing data deluge in crowd sensing, in-network processing is also a promising research direction per se. In fact, current portable devices also own non-negligible computing capabilities, which may be used for distributed processing. First envisioned in [4.62], the idea has been then investigated in [4.63], where its viability is studied and a framework for its implementation is proposed. A first attempt to deploy these concepts is represented by [4.64], while in [4.65][4.66][4.61] the possibility of involving opportunistic users in pervasive environment sensing is explored. We believe that *there is still a lot of research in this direction*: one can envision a connected world of Internet, fixed infrastructures, opportunistic nodes and smart things [4.67][4.68], where users equipped with wireless devices interact seamlessly with all the network components, acting not only as final service users, but also as providers of data, memory, processing and bandwidth resources to others. In this sense, opportunistic networking becomes another enabling technology in the development of the Internet of Things.

**4.4. ONs in HetNets**

Due to the unprecedented mobile traffic growth, not compensated by an equivalent increase in cellular capacity, a research trend of the last few years has been to exploit heterogeneous wireless technologies for cellular traffic offloading [4.69][4.70]. Within this framework, opportunistic device-to-device communications operate in synergy with the cellular infrastructure and are seen as capacity enhancer in a terminal-to-terminal delayed offloading scenario [4.71]. Content dissemination is an application particularly relevant to data offloading, since a single piece of content is shared by many interested users. Thus, once a few copies are injected via the cellular infrastructure, the remaining ones can be disseminated by the users themselves, which communicate opportunistically with each other. The cellular network only takes over in case content has yet to reach the interested users after a predetermined deadline. Research on this topic has pursued two main directions: the optimal selection of the subset of nodes to which the content should be initially injected [4.72][4.73][4.74] and the understanding of what type of traffic should be offloaded and how [4.75][4.76][4.69]. On the technological side, the LTE Direct technology [4.77] is under development, with the specific aim of allowing discovery of resources (e.g. people, services) in the same cell as the device, and direct device-to-device communication. This technology is still in its infancy, research in the field is at the very beginning, and its potential and impact have yet to be determined.

Despite the fervent research activities in the last few years, several problems remain open, such as how to integrate in a coherent architecture the different data offloading opportunities, or to what extent the operator should impose a control on the content dissemination process (see [4.71] for a thorough discussion). Besides these, ON-based data offloading shares many open problems with standalone opportunistic networks, such as willingness of users to cooperate, energy issues, and lack of test beds.

**4.5. Cloud-assisted ONs**

In a heterogeneous network, the availability of a cellular link connecting mobile devices to the Internet opens up the possibility of cloud-assisted networking. The cloud can be exploited to assist either the cellular infrastructure or mobile devices. As far as mobile devices are concerned, several proposals [4.78][4.79][4.80][4.81] suggest offloading heavy computations to the cloud in order to save energy on mobile devices (thus addressing also the energy issues discussed previously). Unfortunately, in practice currently available solutions are able to offload very few computations, and even worse, since the most energy hungry applications are actually those that involve communications, rather than computation, current solutions do not prove very effective [4.80]. An original take on using the cloud infrastructure is discussed in [4.82], where the authors suggest performing neighbour discovery on the cloud (since it only involves sending a tiny amount of information) and delegating data exchange to local pairwise WiFi contacts. When used for offloading the cellular infrastructure, the cloud is usually assigned the role of implementing the control plane of the data offloading process [4.83]. This relieves the burden on the cellular infrastructure while at the same time moving the control of content dissemination from the cellular operators to third-parties, thus potentially opening up new business opportunities. Despite these preliminary contributions, however, the role a cloud infrastructure can play in hybrid opportunistic networks is still largely unexplored and we believe it deserves significantly more attention from the research community.

**4.6 Grand challenges**

Three main directions have characterized the ON research: human mobility & mobility models, routing protocols and data dissemination. The research in these areas is now quite consolidated and there is a very limited space for novel and high-quality contributions. On the other hand, new exciting research opportunities exist in addressing the following topics:

- Make ONs enter people's lives: dealing with this aspect requires delving into several facets, namely, *(i)* understanding how people behave and interact in an ON, and motivating users to join an ON, so as to both design appropriate solutions for real ONs and guarantee sufficient density of users in order to have effective opportunistic communication support; *(ii)* deploying prototypes and applications on a large scale; *(iii)* addressing the technological limitations that make multi-hop ad hoc communications between off-the-shelf devices a daunting task, also paying attention to improving and optimising the energy consumed by the portable devices participating in an ON.
- ONs and IoT: researches should investigate how opportunistic networking can be brought into the crowd sensing/computing and IoT domain.
- ONs and other networking technologies: in order to extend the applicability of ONs, it is important to integrate ONs into a coherent architecture for HetNets and continue exploring the benefits that the integration of heterogeneous networking paradigms can bring.
- ONs and clouds: following recent developments in network architecture, it is interesting to study how ONs can be enriched and assisted by a synergy with a cloud infrastructure, when available.

## 5 Vehicular Networking[6]

Vehicular Ad hoc NETwork (VANET) is a multi-hop ad hoc network made up of vehicles that communicate among them by exploiting wireless technologies, generally, belonging to the 802.11 family. In this network the mobility of the network nodes (i.e., vehicles) is constrained by the road characteristics and the other vehicles moving along the road. Typically, power consumption is not an issue for this network, as vehicle batteries are continuously recharged. The high socio-economic value of vehicular applications pushed the international standardization bodies to develop technical specifications to be adopted by vehicle industry. Among these it is worth reminding the IEEE 1609 standard family for Wireless Access in Vehicular Environment (WAVE) that has been developed upon the IEEE 802.11p standard [5.1].

     VANET is a specialization of the multi-hop ad hoc network paradigm well motivated by the socio-economic value of advanced Intelligent Transportation Systems (ITS) aimed at reducing the traffic congestions, the high number of traffic road accidents, etc. Indeed VANET can support a large plethora of applications including safety traffic applications (e.g., collision avoidance, road obstacle warning, safety message disseminations, etc.), traffic information and infotainment services (e.g., games, multimedia streaming, etc.). For example, a car involved in an accident can exploit the possibility to directly communicate with other vehicles to inform nearby vehicles of the dangerous situation. Advanced ITS systems require both vehicle-to-roadside (V2R) and vehicle-to-vehicle (V2V) communications. In V2R communications a vehicle typically exploits infrastructure-based wireless technologies, such as cellular networks, WiMAX and WiFi, to communicate with a roadside base station/access point. However the roadside units are not dense enough to guarantee the network coverage required by ITS applications, and hence V2V communications are adopted to extend the network connectivity/coverage and to guarantee better network performance. V2V communications are based on the "pure" multi-hop ad hoc network paradigm as MANET. Specifically, according to this paradigm, vehicles on the road dynamically self-organize in a VANET by exploiting their wireless communication interfaces. Therefore, in its initial stages, VANETs were thought of as a specialization of MANETs, where nodes moved along streets. As expected, researchers tried to apply existing MANET solutions (e.g. routing) to VANET scenarios. However, VANETs have very distinctive properties from MANET, and hence require very specific technologies and protocols. In particular, they exhibit a different network architecture, a topology with frequent network partitions, an uneven network density, as well as very strong demand for scalability due to the large number of vehicles that may form a VANET. Thus, it is well-known that MANET (and Wireless Sensor Network -- WSN) solutions simply do not work in VANETs. Nowadays, this is something highly accepted and well-known within the research community. But, from time to time, we still receive contributions based on

---

[6] By Pedro M. Ruiz

MANET or WSN protocols, which clearly do not scale with the number of nodes, or do not take advantage of any of the properties of VANETs such as the constrained mobility, the absence of strong energy constraints, or the ability of nodes to know their GPS position.

The routing task is challenging in VANET due to the high mobility of vehicles and the possibility of sparse networking scenarios, which occur when the traffic intensity is low. This makes inefficient the legacy store-and-forward communication paradigm used in MANET and push toward the adoption of the more flexible, pragmatic and robust store-carry-and-forward paradigm adopted by the opportunistic networks (see the "Opportunistic Networking" section) [5.2]. In addition, whenever possible, V2V communications exploit V2R communications to make their communications more robust reducing some weaknesses and vulnerabilities of pure infrastructure-less communications.

In VANETs, a special attention has been reserved to the development of optimized one-to-all (in a specific region) routing protocols as several applications developed for VANET use broadcasting (geocasting) communication services [5.3] to distribute a message from a vehicle to all other vehicles (in a given area). Broadcasting or geocasting are also the basic communication services for content dissemination in a VANET, i.e., the dissemination to other vehicles of a file containing relevant information (e.g., a city map or infotainment information such as a music mp3 file). Due to the intermittent connectivity conditions, the opportunistic paradigm applied to vehicular networks has recently generated a large body of literature mainly on routing protocols and data dissemination in vehicular networks (e.g., [5.4][5.5]).

In the current research scenario, a practice to avoid is that of proposing minor improvements to an existing protocol in a very specific scenario and network conditions. For example, a minor change to a routing protocol to achieve a limited increase in performance (e.g., only when vehicles are at a traffic jam), cannot be accepted for publication in a high quality journal. In addition, while so far we have published a plethora of communication protocols trying to solve very specific problems in very specific conditions (e.g., MAC, routing, management protocols, etc. [5.6]-[5.9]), now we need to make a step forward, where the new protocol proposals should be integrated in a wider network architecture.

We need to focus our future research in heterogeneous scenarios in which vehicles can exploit and combine all types of possible communication technologies to support advanced ITS applications. The case for communication protocols for hybrid VANETs has finally become the common scenario. During the last five years almost every communications paper in the area of VANETs considers such scenario with Vehicle-to-vehicle (V2V), Vehicle-to-Infrastructure (V2I) communications as well as road-side units (RSU) and access to cellular networks (e.g. LTE.). So, unless there is a very innovative and disruptive solution or technology, specific data communication protocols for VANETs, which cannot be integrated into a wider heterogeneous scenario with a diversity of devices, networks and applications, can be considered a cold topic in VANET research.

A survey of the VANET research is presented, in the chapter by Awwad, Yi and Stojmenovic, in [1.1]. Few other chapters in same book survey the ongoing research activities on experimental testbeds and simulation tools.

**5.1. Research Challenges**

Vehicular Ad Hoc Networks continue being one of the most promising technological developments in the area of computer communications. After more than a decade of research and great advances in the field, interest in vehicular networks has extended beyond academia. ICT companies and car manufactures have come into play and many standardization efforts are also on their way. Moreover, public authorities, such as the European Commission (EC) in Europe and the Department of Transportation (DoT) in the USA, are also getting involved and promoting several *Field Operational Test*. The U.S. Federal Government announced last February that National Highway Traffic Safety Administration (NHTSA) is starting work on a regulatory proposal to require Vehicle-to-Vehicle (V2V) devices in new vehicles, sometime, in the coming years. In Europe, ETSI has finally standardized higher layer networking protocols.

With such a context, one may question whether the remaining efforts are just industrial, or there is still room for academic research. We must admit that many basic VANET communications topics lack research interest nowadays. However, there are open research challenges that need to be solved before VANET can be deployed to its full potential. In this section, we elaborate on that, with the aim of helping journal readers find interesting research opportunities. Hereafter, we discuss, without being exhaustive, some research opportunities in the context of VANETs.

**Knowing fundamental performance limits.** While there are areas in VANET research with a very large number of research contributions, this is one of those in which we are lacking contributions. There are some papers on fundamental limits for data aggregation [5.10], asymptotic throughput capacity [5.11], the optimal broadcasting structure [5.12], and the multicast capacity for hybrid VANETs with directional antennas [5.13]. However, there are still many of the well-known fundamental limits questions that remain un-answered.

**Dynamic Spectrum Access (DSA) for inter-vehicle communications**. Although the U.S. Federal Communications Commission (FCC) has allocated a total of 75MHz in the 5.9 GHz for DSRC, it is a reality that vehicular networks suffer from spectrum scarcity in scenarios with high density of nodes, or when infotainment services using high data rates, are considered. DSA has been one of the recent hot topics in wireless communications and can be a very interesting technology to solve the spectrum scarcity problem in VANETs. The basic idea is allowing unlicensed users to opportunistically exploit the spectrum opportunities that licensed users are not using. For instance, vehicles can opportunistically access to digital TV or 3G/4G licensed bands to get better QoS. Recent works over TV White Space (TVWS) have proven the feasibility of this approach, usually referred as Cognitive Radio VANETs (CR-VANETs) feasible [5.14]. However, traditional DSA schemes assume static users. Thus, using DSA, cognitive radios and cooperative communications in VANETs poses a number of research challenges. During the last four years there has been an increasing number of papers on CR-VANETs, but there are still interesting open issues in many areas of VANETs such as routing, Quality of Experience (QoE), handling mobility and even on security. The reader can refer to [5.15] for a state-of-the-art in the matter and a list of interesting open issues.

**Simulation, validation and experimental results.** Both simulation and experimentation have a major role in designing and evaluating the solutions developed for vehicular networks. It is widely recognized that simulation results are only significant when realistic models are considered within the simulation tool-chain. Indeed, in the simulation of vehicular networks a lot of attention has been dedicated to develop realistic models of roads, and of the vehicle mobility, by exploiting the extensive literature developed in the field of transportation systems, e.g., models of how cars move along a road taking into consideration their speeds, the distance among them, traffic signals, the road layout, etc. [5.16]. However, quite often research works on the subject are based on simplistic models unable to capture the unique characteristics of vehicular communication networks. As stated in [5.17], we count nowadays with enough tools to perform high-fidelity simulations and there is no excuse for not using them. For example, ns-2, ns-3, SWANS, OMNET++, OPNET are able to take as input trace files, generated by specialized vehicle mobility models (e.g., VanetMobiSim and SUMO), which provide realistic vehicle mobility behaviors. Of course, it would be desirable to have higher level of detail but for many of our research needs, they are adequate. However, in some scenarios, the interdependences among vehicle communications and their mobility patterns make the problem more complex. For example, when a congestion/accident occurs the communication among vehicles influences their mobility by triggering a road change and/or a speed reduction. For this kind of simulation the above approach is not suitable and therefore the VANET community is currently developing simulators that are able to take into account the interdependencies between the vehicles' mobility and their communications. For example, TraNS (Traffic and Network Simulation environment) integrates ns-2 and SUMO by providing feedbacks from the network simulator (ns-2) to affect the mobility traces produced by the SUMO mobility simulator. Another important aspect to consider (to produce realistic VANET simulations) is related to the simulation of the wireless channel among vehicles and to/from roadside units taking into account how the radio signal propagates in this environment. An updated and in-depth discussion of VANET simulations models and tools can be found in "Mobility Models, Topology and Simulations in Vanet" [1.1].

At this stage, more research is needed in validating existing simulation software and models compared to experimental data. VANET prototypes have been extensively used to verify the feasibility of VANET solutions and the effectiveness of VANET applications. Several testbeds have been developed worldwide. Among these, it is worth remembering the CarTel project at MIT (which developed a 27-car testbed for testing V2V, V2R, road surface monitoring, etc.), the DiselNet project at UMass (which is based on 35 buses which implement a DTN-based communication paradigm), the ShanghaiGrid project (which involves thousands of taxies and buses for vehicle and traffic monitoring and environment sensing) and the pioneering FleetNet project (involving several companies and university in Germany to demonstrate their platform for inter-vehicle communications). An updated survey of VANET experimental activities can be found in "Experimental work on Vanet" [1.1]. We know the difficulty of running large-scale VANET experiments. However, for many parts of the system, small-scale experiments can serve to validate models and simulations and such research is both needed and valuable.

**Highly Heterogeneous VANET systems.** There have been a number of solutions to deal with very specific problems in hybrid VANETs encompassing V2V, V2I and V2R communications such as broadcasting, data dissemination, routing, etc. However, we still need a lot of research to effectively manage those highly heterogeneous scenarios consisting of a plethora of technologies, addressing schemes, and simultaneously deployed applications. This kind of scenarios where everything is connected to everything using different technologies are difficult to handle and have been referred recently as Internet of Vehicles (IoV). One of the promising research directions in the short term is exploiting cross-layer designs specially for complementing ad hoc and existing cellular technologies so that applications can get the service they need, without having to know the underlying details of where the packet is to be routed. Such intelligent strategies for traffic offloading can help mitigate congestion problems while improving the overall QoS. There are many other mid-term and long-term open research questions in such scenarios such as how to accommodate security without negatively impacting some of the devices (e.g. energy in sensors), how to orchestrate the operation of data communications while still preserving the required scalability constraints, how to manage QoS so that competing applications with strict requirements (e.g. safety and infotainment) can coexist, etc. General directions on how to explore research solutions in these highly heterogeneous scenarios can be found in [5.18].

**Application-aware VANET networking.** One common pitfall that we often see in many VANET papers is that of considering communication scenarios that are not relevant for vehicular network applications. They can be an interesting engineering exercise, but that certainly does not help advancing the field. Application needs are the ones that drive communication protocol design and not the other way around. In WSNs, it is a common practice to employ cross-layer schemes whereby network protocols try to minimize energy consumption by only getting data that is relevant for the applications, or getting only enough level of detail about the data. However, such close interrelation between applications and networking protocols have not been exploited enough in VANETs. With the advent of the so called Vehicular Sensor Networks (VSN) in which vehicles act as moving sensors collecting data, we should explore if VANET networking protocols can employ similar data-centric techniques. Why asking all vehicles in an area to report data? How many is enough to detect traffic congestion? Mobility and other VANET properties make solutions such as compressive sensing and similar schemes not directly applicable to VANETs. However, finding similar data-centric designs for VANETs is certainly an interesting research topic that can bring important benefits in terms of reducing spectrum usage, improving performance, etc.

**Information-centric networking for VANETs**. Significant enhancements in VANET are expected by adapting new communication paradigms to VANETs. For instance, it is worth exploring the use of Information-Centric Networking (ICN) [5.19]-[5.21] in vehicular networks because it can bring important benefits. By using named content, name-based routing, and in-network content caching applications can just say what they need without knowing where it is. Although there exists already some works using ICN [5.22][5.23] and content caching strategies in VANETs [5.24], there are still many open research questions on how to make it work efficiently in such highly heterogeneous scenarios.

**Software Defined VANETs**. As we have learned from the Internet, large-scale deployment requires a very flexible architecture. For the case of VANETs we are in very similar situation where industry is starting to consider some deployments, but getting global acceptance of protocols and being able to change them in the future may not be really easy. One promising direction to avoid repeating the same mistake we did with the Internet could be employing the emerging network paradigm of Software Defined Networking (SDN). The idea is to separate the control and data planes, so that the forwarding behaviour of vehicles can be controlled externally from the underlying network infrastructure. Some initial study in [5.25] established a starting proposal on how SDN can be brought to VANETs. However, in the same way that SDN is still a growing research discipline, its application to VANET is also in its early stages and there is a lot of research to be done. However, this new paradigm can bring interesting benefits so as to make highly heterogeneous VANET systems mentioned above easier to manage and easier to integrate in the future with other new paradigms such as IoT and the so called Internet of Vehicles (IoV).

**Security and privacy**. Security is one of the research areas that have received significant attention from VANET researchers since its inception [5.26][5.27]. Obviously, many of the envisioned VANET applications (specially safety) cannot be really deployed without strong security and privacy guarantees. There has been a significant advance in security and privacy for VANETs during the last decade. In fact, as stated in [5.28], there are aspects such as using Elliptic Curve Cryptography (ECC), Public Key Infrastructures (PKI) and Digital Certificates (DC) in which there is clear agreement. This is also the case for aspects such as guaranteeing message integrity through the use of digital signatures and the use of tamper-proof devices in vehicles. Solving privacy issues using pseudonyms and using reputation schemes to monitor consistency and detect misbehaviour are also highly researched topics in which most research issues are solved. As in MANET, security mechanisms that enforce the correctness and integrity of the network operations are not sufficient and mechanisms to guarantee/enforce the cooperation among network nodes have also been devised (e.g., see [5.29], and the reference herein).

To summarize there are many security aspects for which we have good solutions. However, we still lack scalable solutions being able to work in large-scale deployments. We also lack solutions that take into account how to achieve a good trade-off between security and the amount of required network resources that are needed for its operation. In particular, when we consider highly heterogeneous scenarios as mentioned above, we still lack an integrated and general security and privacy solution that can work with such level of diversity of devices and technologies while still being effective and practical. In addition, we need to further investigate on security of devices and low-level security. As recently shown [5.30], even if protocols are secure, if systems are not, car hacking can become something very common in the near future.

## 6. Wireless Sensor Networks[7]

The increased success in integrating communication and sensing circuitries on miniaturized devices has led to the emergence of the Wireless Sensor Networks (WSNs) paradigm [6.1]-[6.6]. WSNs represent a "special" class of multi-hop ad hoc networks that are developed to control and monitor a wide range of events and phenomena, e.g., precision agriculture, surveillance, pollution monitoring, health monitoring, structural monitoring [6.7]-[6.13]. The application driven view, and the need for solving concrete problems, made WSNs successful both in the academia and industry.

In a WSN, a number of sensor nodes are deployed to monitor an area. These nodes can communicate directly among themselves by exploiting their wireless interfaces, and typically pursue multi-hop paths (through the other sensor nodes) to disseminate the collected information toward a sink node. The sink node interface the WSN to remote command centers through dedicated long-haul links and/or other interested users through the Internet. The design of WSNs highly depend on the specific application scenario and performance requirements in terms of reliability, timeliness etc. For some of these applications, such as space exploration, coastal and border protection, combat field reconnaissance and search and rescue, it is envisioned that a set of mobile sensor nodes will be employed to collaboratively monitor an area of interest and track certain events or phenomena. By deploying these sensors for autonomous/unattended operation in harsh environments, risk to human life can be avoided and the cost of the application can also be lowered. In other applications such as structure monitoring and manufacturing automation, the WSNs operate in controlled setups.

---

[7] By Mohamed Younis

Initially researchers considered energy and resource constraints to be the key challenge in designing WSNs in comparison with general ad-hoc networks [6.14][6.15]. Scalability, security and data fidelity have also been deemed design objectives, which made work on routing, MAC, clustering, coverage, data integrity, etc. to be dominating. A comprehensive library of techniques has been developed over the past 15 years or so. After such extensive research, the technical community expects mature solutions to emerge for practical application scenarios. Consequently cross-layer and multi-objective design methodologies are of great interest. In addition, supporting applications in 3D environments [6.16] and under unstable and dynamic connectivity considerations are also of interest [6.17]. Examples include underwater sensor networks, airborne sensor networks, and robot-assisted sensor networks [6.18]. The following subsections highlight what we view as being accomplished (extensively researched) and what the research community ought to focus on in the next years.

## 6.1 Conventional and Widely-Researched Design Issues

The following areas have received the most attention of the WSN researchers. The results have enriched the literature with numerous publications. We just provide general highlights and point out few topics that are worthy further investigation related to these areas.

- Node Placement: Careful node placement can be a very effective optimization means for achieving the desired design goals, e.g., energy costs, coverage, connectivity [6.19]. Since many variants of the placement optimization problem is NP-hard, numerous heuristics have been proposed. Published work spans stationary and dynamic node positioning schemes, random and controlled placement strategies, as well as analytical studies [6.20].

- Topology Management: The primary objective of existing topology management techniques in WSNs is to achieve sustainable coverage while maintaining network connectivity and conserving energy. For example, these techniques are employed to track the status of communication links among nodes, to conserve energy by switching off some of the nodes without degrading network coverage or connectivity, to support hierarchical task assignment for data aggregation, to balance the load on existing nodes and links, or to provide scalability by minimizing medium access collision and limiting overhead [6.21]-[6.25]. Provisioning multi-tier hierarchy through node grouping into clusters is one of the popular topology formation methodologies with techniques factoring in all sort of parameters and metrics, e.g., resource conservation, load, energy harvesting, delay, etc. [6.26][6.27].

- Medium Access Arbitration: The distributed operation of typical WSNs has fuelled interest in devising MAC protocols that are lightweight in terms of control messaging overhead, limits passive listening and support sleep modes [6.28]-[6.30]. While quite a few of the published MAC protocols could yield good performance using single medium access sharing methodologies, e.g., time or frequency base multiple access, a number of protocols, such as [6.31][6.32] have demonstrated that combing multiple methodologies is more effective. Some of the protocols have matured and become widely accepted as the de-facto standard [6.33]. Support for quality of service and exploiting advanced physical layer features have been also be studied as well [6.34]-[6.36].

- Routing, data dissemination and fusion: Routing data in WSNs is an area that has been researched quite extensively, or probably exhaustively. Both flat and hierarchical routing trees have been investigated. Numerous centralized and distributed path formation schemes have been developed [6.37]-[6.39]. In the design of a WSN routing protocol, energy efficiency is a main design objective [6.40][6.41]. Routing schemes with mobile nodes and/or mobile sinks have been also extensively investigated [6.42]-[6.45]. Data aggregation and fusion have been factored in some of the work [6.46]-[6.51]. Link and network cross-layer optimization has been popular as well [6.52].

- Secure Communication: The resource constraints of the sensor nodes and the lack of continuously-accessible trusted authorities have been deemed the major differentiator for WSNs from contemporary networks [6.53][6.54][6.55]. One of the research thrusts has been in the management of cryptographic keys [6.56]. Nonetheless, the bulk of the research efforts have been dedicated to devising lightweight versions of contemporary security solutions in order to suit WSNs and trading off attack resilience for resource conservation [6.57]-[6.60].

After the development of an extensive library of techniques, the community is eager to see systems that utilize a subset of these techniques in the implementation of real life applications. It is argued that it is time to judge the practicality of what has been developed in the above areas of research. Therefore, experience with engineered systems and actual WSN deployment is of great interest; particularly cross-layer optimization and synergetic interaction among the various techniques will be invaluable for both researchers and practitioners. In addition, some specific problems can still benefit from additional research efforts despite the major progress made to-date. The following topics can be noted:

- Decentralized node placement and topology self-configuration: While quite a few techniques have pursued self-spreading of nodes and relocation of assets, adaptive node positioning solutions that can deal with dynamic changes in the application requirements and emerging events are of interest. Example scenarios include the autonomous recovery from failure and self-reconfiguration to cope with varying QoS requirements.  In many WSN applications, the network is deployed in harsh environments such as battlefields where the nodes are susceptible to damage. Although a number of distributed solutions have been proposed for tolerating single node failures [6.61], the loss of multiple collocated nodes is very challenging. Distributed recovery schemes for such multi-node failure are in demand. A similar scenario is when QoS requirements, e.g., coverage and data fidelity, dynamically vary due to the emergence of an event that warrants reshaping of the WSN topology.

- Anti-traffic analysis measures: Intercepting a wireless transmission enables an adversary to locate the sender and estimate its proximity to the receiver. Analyzing the WSN traffic by correlating interceptions throughout the network enables the adversary to locate key network assets, namely, the base-station and active data sources, which may then be targeted by pinpointed attacks. This is serious vulnerability that cannot be mitigated by contemporary security primitives such as packet encryption. Similarly, camouflaging or hiding nodes does not provide protection when their location is unveiled. Furthermore, employing spread spectrum technologies does not constitute a sufficient countermeasure since it reduces rather than eliminates the prospect of transmission detection and interception. Moreover, the broadcast nature of the wireless links and the traffic pattern in the network make anonymous communication techniques developed in the realm of peer-to-peer networks inapplicable and ineffective for concealing the location of WSN nodes. While location privacy of data sources in the realm of WSN has received some attention [6.62], we argue that in many application scenarios the sources are not as important given the implicit node redundancy. Instead, the base-station becomes the favorite target for the adversary, especially when considering its major role in managing the WSN and interfacing it with remote command centers [6.63][6.64][6.65]. Therefore, we view the problem of guarding a base-station against traffic analysis attacks warrants increased attention.

- WSN with voluminous data: In applications that involve imaging and multi-media sensors [6.66], the volume of data overwhelms the network communication resources and in-network data processing becomes necessary. Of particular interest is the integration of emerging techniques for compressive sensing and big data analytics with contemporary WSN management primitives [6.67]. Computation and communication trade-off, balancing delay, energy and data fidelity metrics, and the pursuance of heterogeneity in WSN design are important issues that are worthy of investigation.

**6.2 Research topics of growing importance**

In recent years a number of emerging WSN applications have stimulated interest and motivated researchers to refocus their efforts.  We have grouped the areas that we believe are worthy attention into two sets. The following enumerates these areas, briefly highlights key issues and points out publications that can be checked for detailed analysis. Our objective is to hopefully steer the efforts towards these areas and direct the readers to open research problems in WSNs.

*6.2.1 WSN for Challenging Environments/Applications*

With the continual advance in low-power and high-density electronics and increased level of miniaturization, WSNs have being gaining ground and finding applications in challenging environments and need to achieve very ambitious objectives. The following highlights popular examples of these applications:

- Underwater WSNs: Examples of underwater WSN applications include environmental state monitoring, search and rescue, oceanic profile measurements, leak detection in oil fields, tracking seismic activities, seabed profiling, search-and-rescue and coastal patrol missions, distributed surveillance, and navigation [6.68]-[6.74]. For these applications, either a set of sensor nodes or autonomous underwater vehicles (AUVs) will be deployed to collaboratively monitor an area of interest and track targets, e.g., submarines, or track certain phenomena. Since radio waves tend to get absorbed in water, underwater applications have to rely on free-space optics [6.75], laser [6.76], or acoustic [6.77]-[6.79] rather than radio channels. However, free-space optics have a limited range, using laser requires significantly high power, and acoustic channels are slow and sensitive to the water properties, suffer major distortion, and have low bandwidth. To be effective, directional transmission will be required or at least highly recommended for these channel types. Therefore, the link establishment will be further complicated by the inherit mobility of the node due to the water current. Furthermore, the mobility, link instability and the 3-D environment make node localization and topology management very challenging. Although some progress has been made over the past few years and node platforms and products have started to emerge [6.69][6.80][6.81], the research community has a way to go before mature solutions are developed. For a detailed discussion of the design issues of underwater sensor networks, the reader is referred to [6.82]-[6.86].

- Underground WSNs: In most developed societies, utility companies rely on an extensive network of pipelines for supplying natural gas and water to residences and industries. The bulk of the utility pipeline system is laid belowground and would significantly benefit from the use of WSN to monitor the structural integrity of the pipe segments, and to detect and locate leaks and clogs. Conserving water and countering pollution due to hazardous leaks have been a focal point for the public and government agencies; thus detecting and containing problems in early stages would be invaluable. Underground WSNs would also be useful for boarder security, e.g., detecting infiltration through illegally dug tunnels, and for tactical military missions, e.g., the deployment of smart landmines. In addition, underground WSNs find applications in precision agriculture by tracking the moisture and minerals levels in soils, and in disaster warning by monitoring seismic activities for early landslides and earthquakes. However, the implementation of underground WSNs faces major challenges [6.87][6.88]. Unlike aboveground deployment, the propagation of electromagnetic signals is significantly restrained with a loss of more than 100 dB per meter [6.89]. Such high signal attenuation varies based on the type of soil and the moisture level. Raising the antenna of the individual nodes above ground and the use of wired links is logistically infeasible or impractical in most applications. While the topology of an underground WSN is controlled, the establishment of reliable communication links and harvesting energy for powering up the nodes complicate the node placement optimization. Few solutions have been worked out so far [6.90]-[6.94], and major research efforts are needed to tackle the underground WSNs design issues [6.95].

- Body Area WSNs: A body sensor network fundamentally consists of a set of wearable and implanted devices in the human body to continuously assess the health status through measuring key indicators, e.g., blood pressure, heart beat, etc. Meanwhile, the notion of Body Area Network (BAN) is a bit broader and involves devices that are not mounted on the human body. Although a BAN is widely-known as a tool for conducting medical studies and as an outpatient healthcare instrument [6.96]-[6.98], applications in the entertaining and gaming industry have recently emerged [2.16]. Most of the technical challenges for BANs are motivated by the physical characteristics of the involved devices. Fundamentally most of the numerous ongoing projects are multidisciplinary in nature [2.16][6.96]. In addition to the complexity of the application-specific data fusion, the main design challenges include coping with the very stringent energy constraints for the employed devices and the relatively high path-loss when transmitting at low-power as a means to extend the network lifetime. These challenges become even more prominent with emergence of nano-scale devices [6.99][6.100]. A major push from the biomedical industry to standardize the protocol stack and the application interface has yielded the IEEE 802.15.6 standard. However, the IEEE 802.15.6 standard opts to ensure interoperability rather than providing complete specifications of the MAC protocol. Therefore, optimal medium access sharing among the BAN nodes is an open and complex research problem given the energy constraints, delay sensitivity of the transmitted data, and the signal interface that the miniaturized BAN devices are subject to. In addition, supporting the integration of non-wearable sensors is an interesting research direction. With technological advances in gesture and facial expression analysis, a BAN can grow in scope to include sensor data that reflects the appearance of a human and factor that into the overall assessment of the body condition [6.101][6.102]. For example, the presence of back pain or distress can be inferred if someone is bending during a walk.

- Airborne WSNs: The recent years have witnessed major growth in the popularity and availability of small and battery-operated unmanned airborne platforms such as quadcopters and drones. The reliability and capabilities of these platforms have reached such an advanced level that an online retailer like Amazon is considering the use of quadcopters for package delivery if it receives regulatory approval [6.103]. The size of some of these platforms is as small as that of a bird and can be equipped with a wide variety of sensors. While unattended airborne vehicles (UAVs) have been employed in some WSNs as base-stations and as relays in order to enable connectivity among sensors on the ground [6.104][6.105][6.106], research on the use of a group of networked UAVs as an airborne WSN is still in an early stage. Compared to ground-based WSNs, a networked UAVs are not constrained by terrain and enable wide field-of-view sensing. In addition, an airborne WSN provides high levels of agility and adaptability to environment changes, and achieve broad temporal and spatial coverage. Given these advantages, the interest in the use of airborne WSNs in scientific, defense and civil applications has been rising [6.107]-[6.112]. Nonetheless, the networking of UAVs is quite challenging, and mature and practical solutions for many of the design issues are yet to be developed. Most notable among the technical challenges are achieving accurate node localization, sustaining network connectivity, and supporting coordinated navigation and coverage. While airborne WSNs resemble MANETs in terms of mobility and energy constraints, the 3-D nature and the speed of node motion introduce major complications that would render published solutions for topology management in the realm of MANETs ineffective. In addition, the small form factor, weight and battery capacity constraints may make it infeasible to equip a node with GPS receivers; therefore, node localization and navigation have to be based on relative rather absolute coordinates and conducted in a cooperative manner while relying on error-prone techniques such as accelerometers and radio ranging [6.113][6.114].

*6.2.2 Federated Sensory Systems*

The conventional view of a WSN is that it is an application specific. This view still holds and is consistent with the cost, energy and size constraints that a node is usually subject to, and which hinder the provision of generic capabilities. Nonetheless, with the increased deployment of WSNs it is envisioned that data sharing and sporadic and/or opportunistic interactions among multiple WSNs will be necessary to achieve emerging application needs. For example, the BANs of two individuals may exchange alerts when they get in range of one another. These alerts may be on pollution and unhealthy conditions or the risk that one of the two individuals may pose to the other, e.g., when some infection is suspected due to a high temperature. We view federation of multiple WSNs to be a major trend in the future where the capabilities of the individual WSNs will be combined to form a networked sensory system. Such federation is characterized as short-lived integration of features, e.g., sensing modality, and leverage of services (target tracking). In the following we highlight some federation scenarios and points out some technical issues that arise. Among federated sensory systems, the widespread use of mobile and other personal devices is generating the emergence of a new paradigm for sensing the physical world with human in the loop: *phone sensing*. This paradigm is generating a people-centric revolution of wireless sensor networks and therefore it will be discussed in a separate section (see Section 7).

- Inter-networking diverse WSNs: In applications such as disaster management, fugitive tracking, target tracking, and military reconnaissance, it will be desirable to augment the capability of a WSN when some event of interest is detected. The objective for such capability augmentation is to boost the fidelity of the data, engage different and/or more powerful sensing modality, or to cope with increase in the scope/scale of the detected event. While the complexity for achieving such federation varies widely based on the application, three issues may be commonly faced. First, the federated networks may operate under different protocols; therefore, gateways will be required for inter-WSN data exchange. Second, a model will be needed to coordinate the various activities and synchronize the execution of tasks across individual WSNs in order to achieve the federation objective. Thirdly, a temporal and spatial baseline has to be defined to synchronize the time and map the coordinate systems of the federated networks. Only little efforts have been dedicated to the internetworking of WSNs, with the focus on federating an airborne WSN with ground-based [6.114]-[6.118] and with sea-based networks [6.119].

- Space SensorWeb: The vision of the SensorWeb project at NASA is to facilitate the federation of satellite-based sensing capabilities with ground and airborne WSNs [6.120][6.121]. The objective is to support Earth science through multi-modality space- and ground-based sensory systems and to enable access to such a comprehensive sensory system via the Internet. For example, offering data from a diverse set of sensors would improve the understanding of physical phenomena, such as volcanic eruptions, desertification, climate change, tsunami, etc. Currently, a web interface is provided for scheduling satellite observations and a data log is then sent to requesters to be visualized using tools like Google Earth and alike. However, rather than such primitive arrangements, the technical goal is to automate the process so that the space, airborne and ground WSNs become inter-networked and autonomously trigger the observation activities based on the detected phenomena. The realization of the networking aspect of the Space SensorWeb is yet to gear up and many research problems are to be tackled. In addition to the issues mentioned above for inter-networking diverse WSNs, the Space SensorWeb requires the handling of very slow and intermittent satellite links, routing subject to QoS requirements, and dealing with diverse security and access control policies for individual WSNs. In general, the realization of Space SensorWeb involves complex integration issues across many domains such as security, delay tolerant networking, satellite communication, WSNs, etc.

- WSN support for IoT: The prevalence of wireless communication technologies has motivated the introduction of the notion of Internet-of-Things (IoT) [2.2]. An IoT refers to the pervasive interaction between entities, human, objects, animals, using wireless communication links without human coordination. The pervasive and autonomous inter-node interaction within an IoT requires standardized protocol interface. 6LoWPAN is gaining popularity as one of the suitable choices for standardization. Another characteristic of IoT is the lack of stable topology where links typically stay active for short durations due to node mobility. Interfacing WSN with an IoT would lead to the formation of a federated sensory system that serves a wide range of applications such as smart homes, efficient work space, factory automation, etc [6.122]. Basically, the nodes in an IoT can leverage the WSN services without knowing how the network is architected and managed. To illustrate in a smart home one's cell phone or vehicle can communicate with a controller at the garage door to turn on the light when arriving. The controller can connect with a WSN that tracks the weather in the surroundings to estimate the effect of clouds during the day and determine the intensity of light that suits the owner's need while conserving energy. Even such a simple scenario requires major software and protocol support. Complex application scenarios involving mobility of the WSN as well as the IoT nodes is quite challenging [6.123][6.124]. Research on WSN support for IoT is in early stage and is expect to attract lots of attention over the next few years, especially with advances in the development of nanoscale devices [6.125].

## 7. Mobile-phone Sensing[8]

In recent times, mobile phones (and other smart personal electronic devices) have shown rapid improvements in processing power, storage capacities and network data rates, and are quickly growing in the sophistication of the embedded-sensors. The mobile phones of today have evolved from merely being phones to full-fledged computing, sensing and communication devices. It is thus not surprising that over 5 billion people globally have access to mobile phones. These advances in mobile phone technology coupled with their ubiquity have paved the wave for the emerging of people-centric networking (and computing) paradigms. In fact, the major innovation introduced by mobile phones, is the fact that they do move, and move with people (their owners). For this reason, the networking paradigms based on mobile phones are shifting from device-to-device paradigm toward people-to-people paradigm [1.4]. Opportunistic networks have been a first example of this new paradigm in which data forwarding is strongly coupled with the human mobility [7.1], but it is with the phone-sensing paradigm that we fully realize the people-centric networking and computing paradigm [7.2]. In mobile phone sensing (also known as *crowd sensing*) the physical world can be sensed, without deploying a sensor network, by exploiting the billions of users' mobile devices/phones as location-aware data collection instruments for real world observations. The key idea is to empower the citizens' mobile phones to collect and share sensed data from their surrounding environments using their mobile phones [7.2]-[7.5]. Mobile phones, though not built specifically for sensing, can in fact readily function as sophisticated sensors. The camera on mobile phones can be used as video and image sensors. The microphone on the mobile phone, when it is not used for voice conversations, can double up as an acoustic sensor. The embedded GPS receivers on the phone can provide location information. Other embedded sensors such as gyroscopes, accelerometers and proximity sensors can collectively be used to estimate useful contextual information (e.g., is the user walking or traveling on a bicycle). Further, additional sensors can be easily interfaced with the phone via Bluetooth or wired connections, e.g., air pollution or biometric sensors.

In mobile-phone sensing, citizens have a (passive or active) role in sensing the physical world. We typically refer to *participatory sensing* when citizens take an active role in using their devices to sense the world and publish and share the collected information. On the other hand, in *opportunistic sensing* the sensing activities are performed by the (opportunistic) exploitation of all the sensing devices available in the environment, including those available on personal devices without the need of a direct involvement of the devices' owner [7.6].

Participatory sensing offers a number of advantages over traditional sensor networks that entails deploying a large number of static wireless sensor devices, particularly in urban areas. Firstly, since participatory sensing leverages existing sensing (mobile phones) and communication (cellular or WiFi) infrastructure, the deployment costs are virtually zero. Second, the inherent mobility of the phone carriers provides unprecedented spatiotemporal coverage and also makes it possible to observe unpredictable events (which may be excluded by static deployments). Third, using mobile phones as sensors intrinsically affords economies of scale. Fourth, the widespread availability of software development tools for mobile phone platforms and established distribution channels in the form of App stores makes application development and deployment relatively easy. Finally, by including people in the sensing loop, it is now possible to design applications that can dramatically improve the day-to-day lives of individuals and communities. This generates a cyber-physical convergence as the mobile phone acts as proxy of human behavior in the cyber world. While people really move into the physical dimension, their mobile phones sense the physical world and transfer the sensed information in the cyber world. A plethora of novel and exciting participatory sensing applications have emerged in recent years. CarTel [7.7] is a system that uses mobile phones carried in vehicles to collect information about traffic, quality of en-route WiFi access points, and potholes on the road. Micro-Blog [7.8] is an architecture that allows users to share multimedia blogs enhanced with inputs from other physical sensors of the mobile phone. Other applications of participatory sensing include the collection and sharing of information about urban air and noise pollution [7.9][7.10] cyclist experiences [7.11], diets [7.12], or consumer pricing information in offline markets [7.13].

A typical participatory sensing application operates in a centralized fashion, i.e., the sensor data collected by the phones of volunteers are reported (using wireless data communications) to a central server for processing. The sensing tasks on the phones can be triggered manually, automatically or based on the current context. On the server, the data are analyzed and made available in various forms, such as graphical representations or maps showing the sensing results at individual and/or community scale. Simultaneously, the results may be displayed locally on the carriers' mobile phones or accessed by the larger public through web.

---

[8] By Salil Kanhere

Participatory sensing can leverage and yet complicate the operation of and interaction with legacy WSNs. Basically, voluntary participants can augment the capabilities of a WSN and provide a wealth of additional sensor data. Despite the obvious advantage of increased fidelity through the sensor readings, participatory sensing can also allow bogus and inaccurate data that is a detriment to the quality and trustworthiness of WSN services, particularly when the volume of bogus participant samples is large. In addition, interaction with voluntary participants both in coordination and data collection would be stressful for the WSN communication and energy resources, especially when in-network data processing is applied. Therefore, algorithms are needed to determine when a WSN accepts/solicits participatory sensing based on a careful trade-off between performance and load. Moreover, a suite of protocols needs to be developed to enable participant recruitment, WSN access control, query processing, etc. Furthermore, participatory sensing will introduce privacy issues that WSNs typically are not concerned about [7.14][7.15].

### 7.1 Research Challenges

Since participatory sensing relies on existing networking and sensing infrastructure, the key challenges are associated with handing the data, i.e., data analysis and processing, ensuring data quality, protecting privacy, etc. We now present a short overview of the key research challenges posed by this new and exciting paradigm. We also discuss some recent work in overcoming these challenges.

**Dealing with Incomplete Samples.** Since a participatory sensing system relies on volunteers contributing noise pollution measurements, these measurements can only come from the place and time where the volunteers are present. Furthermore, volunteers may prioritize the use of their mobile devices for other tasks. Or they may choose to collect data only when the phone has sufficient energy. Consequently, samples collected from mobile phones are typically randomly distributed in space and time, and are incomplete. The challenge is thus to recover the original spatiotemporal profile of the phenomenon being monitored from random and incomplete samples obtained via crowdsourcing. In [7.10], the authors have developed an innovative approach to deal with this issue using the technique of compressive sensing, in the context of a noise monitoring application.

**Inferring User Context and Activities.** Inferring the surrounding context (e.g., is the user in a party or a quiet room) and activities (is the user walking, running, traveling by car, etc.) undertaken by the phone carrier is of interest in various participatory sensing applications. This is usually achieved using data collected by one or more embedded sensors in the phone which include accelerometers, microphone, GPS, etc. At the high-level, this is essentially a machine learning problem, since we want to the phones to be embedded with the smarts to recognise human activity. The typical approach adopted is to make use of supervised learning. This involves the following three steps. The first step is to collect properly labelled sensor data for the various categories of interest (i.e. training). The second step involves identifying important features from this training data, which can uniquely identify each category (i.e. fingerprints). The final step is to choose an appropriate classification algorithm (e.g., support vector machines, neural networks, etc.), which can be used to achieve accurate classification.

**Preserving User Privacy.** Current participatory sensing applications are primarily focused on the collection of data on a large scale. Without any suitable protection mechanism however, the mobile phones are transformed into miniature spies, possibly revealing private information about their owners. Possible intrusions into a user's privacy include the recording of intimate discussions, taking photographs of private scenes, or tracing a user's path and monitoring the locations he has visited. Many users are aware of the possible consequences, and may therefore be reluctant to contribute to the sensing campaigns. Since participatory sensing exclusively depends on user-provided data, a high number of participants are required. The users' reluctance to contribute would diminish the impact and relevance of sensing campaigns deployed at large scale, as well as limiting the benefits to the users. To encounter the risk that a user's privacy might be compromised, mechanisms to preserve user privacy are mandatory. The authors in [7.16] present a comprehensive application independent architecture for anonymous tasking and reporting. The infrastructure enables applications to task a mobile device using a new tasking language, anonymously distribute tasks to mobile devices and collect anonymous yet verifiable reports from the devices.

**Evaluating Trustworthiness of Data.** The success of participatory sensing applications hinges on high level of participation from voluntary users. Unfortunately, the very openness which allows anyone to contribute data, also exposes the applications to erroneous and malicious contributions. For instance, users may inadvertently position their devices such that incorrect measurements are recorded, e.g., storing the phone in a bag while being tasked to acquire urban noise information. Malicious users may deliberately pollute sensor data for their own benefits, e.g., a leasing agent may intentionally contribute fabricated low noise readings to promote the properties in a particular suburb. Without confidence in the contributions uploaded by volunteers, the resulting summary statistics will be of little use to the user community. Thus, it is imperative that the application server can evaluate the trustworthiness of contributing devices so that corrupted and malicious contributions are identified. Recent work [7.17] proposes a novel reputation system that employs the Gompertz function for computing device reputation score as a reflection of the trustworthiness of the contributed data.

**Motivating Participants.** For a participatory sensing system to be a success, one of the key challenges is the recruitment of sufficient well-suited participants, those who are suitable for tasks that require domain-specific knowledge or expertise. Without adequate motivation, participants may not be willing enough to contribute, which in turn, reduces the data reliability. An effective approach to address this challenge and provide long-term participation is awarding incentives. Incentives can be broadly classified as intrinsic (such as users participating in a sensing campaign that will improve their community) or extrinsic (such as the system providing compensation for user participation) [7.18]. It is yet unclear which approach is best suited for participatory sensing.

**Conserving Energy.** Note that, the primary usage of mobile phones should be reserved for the users' regular activities such as making calls, Internet access, etc. Users will only volunteer to contribute data if this process does not use up significant battery so as to prevent them from accessing their usual services. Even though, most users charge their phones on a daily basis, it is thus important that participatory sensing applications do not introduce significant energy costs for users. Energy is consumed in all aspects of participatory applications ranging from sensing, processing and data transmission. In particular, some sensors such as GPS consume significantly more energy that others. As such, it is important for participatory applications to make use of these sensors in a conservative manner. In [7.8], the authors present an adaptive scheme for obtaining phone location by switching between the accurate but energy-expensive GPS probing to energy-efficient but less accurate WiFi/cellular localization. Similar approaches can be employed for duty-cycling other energy-hungry sensors.

## 8. Summary and Conclusions

Mobile/multi-hop wireless networking is a mature networking paradigm that provides well-known benefits such as network coverage enlargement, throughput increase by means of spatial reuse, and improved reliability through multi-path availability.

In this article, we have review the current status of the mobile/multi-hop ad hoc networking research. We observed that after two decades of research activities the research on MANET protocols is now a cold research area. Then, after presenting the evolution of wireless technologies supporting multi-hop ad hoc networking, we have discussed the milestones and challenges in mesh, opportunistic, vehicular, and sensor networks. For each paradigm we have pointed out the research topics that we believe are worthy of further attention. We have also discussed the evolution of sensor networks towards the people-centric sensing. Thanks to the increasing diffusion of the smartphones, the people-centric paradigm combines wireless communications and sensor networks with the daily life and behaviors of people. The next step is to exploit the mobile phones and other personal devices to provide localized computing and communication services that are tightly coupled with people and their devices. Specifically, the mobile smart devices, by pooling their resources can start offer services as a *mobile cloud* bringing services and resources closer to where they are needed [1.4][8.1][8.2], thereby avoiding the energy and bandwidth costs associated with accessing the services in the cloud [8.3]. People-centric computing and communication paradigm highly depends on human mobility and social relationships. After extensive studies to characterize the properties of the human mobility and define mobility models, e.g., [4.12] [8.4], the research community is currently investigating the impact of people social behavior on people-centric paradigms, e.g., [8.5]. Indeed, mobile devices act as proxies of the humans in the cyber world and thus inherit the social links of their owners [8.6]. Therefore there is a growing interest in connecting the networking research with human-science research on human social network [8.7][8.8].

It can be expected that, in the near future, the various ad hoc paradigms will be mixed together and/or integrated with infrastructure-based networks, thus generating novel networking paradigms merging the self-* features properties of multi-hop ad hoc networks and the reliability and robustness features of infrastructure based networks [1.4]. Examples of this trend are the hybrid cellular-opportunistic networking, which aims to mitigate the cellular data traffic problems by offloading some traffic by opportunistically exploiting the users' devices (see Section 4), and the vehicular networks where opportunistic networking techniques are exploited to overcome the coverage limitations of roadside infrastructure-based networks (see Section 5). The mix of technologies/paradigms needs, first of all, solutions for supporting seamless mobility among a wide range of mobile access networks [8.9].

We conclude by observing that, several hot/emerging networking paradigms which are affecting infrastructure-based networks, such as [8.10] *Dynamic Spectrum Access* (DSA)/cognitive radio networks [8.11], *Information Centric Networking* (ICN) [8.12], and *Software Defined Networking* (SDN) [8.13], are opening new research directions also in the mobile/multi-hop ad hoc network field, e.g., [8.14]-[8.18].

**References Section 3**

## References Section 4

**References Section 5**

**References Section 6**

**References of Section 7**

## References Section 8